\documentclass[twocolumn]{aastex61}
\pdfoutput=1
\usepackage{epstopdf}
\usepackage{wasysym}
\usepackage{float}
\usepackage{textcomp}
\usepackage{braket}
\usepackage{csquotes}
\usepackage{natbib}

\newcommand{\neowise}{\textit{NEOWISE}}
\newcommand{\COprod}{$Q_{\textrm{\scriptsize CO}_{2}}$}
\newcommand{\wise}{\textit{WISE}}
\newcommand{\coafrho}{$\log_{10} Q_{\textrm{CO}_2}/Af\rho$ }
\newcommand{\Ndust}{$N_\textrm{\scriptsize g}$}

\received{October 1, 2017}
\revised{January 30, 2018}
\accepted{February 19, 2018}

\shorttitle{Behavioral Characteristics and CO+CO$_2$ Production Rates of HTCs}
\shortauthors{Rosser et al.}

\begin{document}

\title{Behavioral Characteristics and CO+CO$_2$ Production Rates of Halley-Type Comets Observed by \textit{NEOWISE}}

\correspondingauthor{Joshua Rosser}
\email{jrosser2@u.rochester.edu}

\author{J. D. Rosser}
\affiliation{University of Rochester
206 Bausch \& Lomb Hall
P.O. Box 270171
Rochester, NY 14627-0171, USA}
\affiliation{Jet Propulsion Laboratory, California Institute of Technology,
4800 Oak Grove Drive,
MS 183-401, Pasadena, CA 91109, USA}

\author{J. M. Bauer}
\affiliation{Jet Propulsion Laboratory, California Institute of Technology,
4800 Oak Grove Drive,
MS 183-401, Pasadena, CA 91109, USA}
\affiliation{Infrared Processing and Analysis Center,
California Institute of Technology,
Pasadena, CA 91109, USA}
\affiliation{ Dept. of Astronomy, University of Maryland,
College Park, MD 20742, USA}

\author{A. K. Mainzer}
\affiliation{Jet Propulsion Laboratory, California Institute of Technology,
4800 Oak Grove Drive,
MS 183-401, Pasadena, CA 91109, USA}

\author{E. Kramer}
\affiliation{Jet Propulsion Laboratory, California Institute of Technology,
4800 Oak Grove Drive,
MS 183-401, Pasadena, CA 91109, USA}

\author{J. R. Masiero}
\affiliation{Jet Propulsion Laboratory, California Institute of Technology,
4800 Oak Grove Drive,
MS 183-401, Pasadena, CA 91109, USA}

\author{C. R. Nugent}
\affiliation{Infrared Processing and Analysis Center,
California Institute of Technology,
Pasadena, CA 91109, USA}

\author{S. Sonnett}
\affiliation{Jet Propulsion Laboratory, California Institute of Technology,
4800 Oak Grove Drive,
MS 183-401, Pasadena, CA 91109, USA}

\author{Y. R. Fern\'andez}
\affiliation{Dept. of Physics, Univ. of Central Florida,
4000 Central Florida Blvd., Orlando FL 32816-2385, USA}

\author{K. Ruecker}
\affiliation{Infrared Processing and Analysis Center,
California Institute of Technology,
Pasadena, CA 91109, USA}

\author{P. Krings}
\affiliation{Infrared Processing and Analysis Center,
California Institute of Technology,
Pasadena, CA 91109, USA}

\author{E. L. Wright}
\affiliation{Department of Physics and Astronomy, University of California,
Los Angeles, CA 90095, USA}

\collaboration{The \wise{} and \neowise{} Teams}

\begin{abstract}

    From the entire dataset of comets observed by \neowise{}, we have analyzed 11 different Halley-Type Comets (HTCs) for dust production rates, CO+CO$_2$ production rates, and nucleus sizes. Incorporating HTCs from previous studies and multiple comet visits we have a total of 21 stacked visits, 13 of which are active and 8 for which we calculated upper limits of production. We determined the nucleus sizes of 27P, P/2006 HR30, P/2012 NJ, and C/2016 S1. Furthermore, we analyzed the relationships between dust production and heliocentric distance, and gas production and heliocentric distance. We concluded that for this population of HTCs, ranging in heliocentric distance from 1.21 AU to 2.66 AU, there was no significant correlation between dust production and heliocentric distance, nor gas production and heliocentric distance.

\end{abstract}


\section{Introduction} \label{sec:intro}

\par Comets are an accessible population of solar system bodies that manifest substantial volatile reservoirs which represent the primordial chemistry of the solar system. Comets are either brief visitors or relative new-comers to the inner solar system and so have spent most of their existence in the “deep freeze” of the outer solar system, where volatiles are relatively undepleted. It is generally inferred that the combined populations that constitute comet reservoirs, for example the Oort Cloud and Kuiper Belt Object populations, are more numerous than the other small body populations for the sizes of a few kilometers (cf. \citealt{2015SSRv..197..191D}, \citealt{bauer2017}). Cometary nuclei are defined as a combination of refractory and volatile materials, however, cometary populations differ by some of the most basic compositional, dynamical, and physical properties, indicating some evolution must take place. Mass loss (cf. \citealt{2014koa..prop..445J}), and selective depletion of specific volatile species like CO \citep{2004come.book..317M,2012LPICo1667.6199M,2011LPI....42.2516A}, are manifestations of these evolutionary effects. Ensemble properties of populations also show evidence of evolution \citep{2013Icar..226.1138F,2004Icar..170..463M,bauer2017} which may affect intermediary \citep{Bauer2013} or end states \citep{2016A&A...585A...9L}. Nonetheless, the prevalence of particular volatiles, such as CO and CO$_2$, may be more common among particular dynamical populations, e.g. long-period comets (LPCs) \citep{2010ApJ...717L..66O,2013Icar..226..777R}. It should be noted, however, that such volatiles are not universally present in detectable quantities for all LPCs. Hence, a component of compositional variation as well as evolutionary effects must be present in the cometary sub-populations. One possible way of discerning how strong compositional variations of particular species, such as CO and CO$_2$, are among LPCs is to study a population that is an evolutionary intermediary to the ultimate demise of the populations' members. Such an intermediary state for the LPCs are the Halley-type comets (HTCs), a subset of short period comets with orbital periods ranging from 20 yrs. to 200 yrs. HTCs have been shown to evolve primarily from LPC reservoirs (\citealt{1999Icar..137...84W}), hence, if HTCs and LPCs share similar origins, both evolving from Oort Cloud populations, HTCs may be used to determine the effects of insolation on LPCs, separate from original composition.

\par Various studies on organics for certain HTCs, e.g. by \citet{8ptuttle} have been completed in the past. However, no previous study has presented a uniform survey of CO+CO$_2$ production for multiple HTCs. This is due primarily to the rarity of space-based platforms that are capable of detecting CO$_2$. Such platforms also detect CO more easily than ground-based telescopes, since they are unencumbered by the Earth’s atmospheric absorption. Observations of HTCs at large Earth-Sun distances is still relatively rare, thus these are some of the first in-depth analyses of CO and CO$_2$ production by such objects.

\par On 2009 December 14, the \textit{Wide-field Infrared Survey Explorer} (\wise{}) was launched to complete a mid-infrared survey of the entire sky at 3.4, 4.6, 12, and 22 $\mu$m. These bands are respectively referred to as the W1, W2, W3, and W4 bands \citep{Wright2010}. The primary cryogen tank was fully exhausted by 2010 August 5, making the W4 band inoperable: this initiated the 3-Band phase (W1, W2, and W3 bands) of the mission until the secondary cryogen tank was depleted on 2010 October 1 \citep{Bauer2015,Mainzer2011c}. After October 1, only the W1 and W2 bands remained operative. The mission then entered the four month \neowise{} Post-Cryogenic Phase with the purpose of finding minor planets, until 2011 February when the telescope went into hibernation \citep{Mainzer2011c,Mainzer2012,Masiero2012}. On 2013 October 3, the \wise{} spacecraft was brought out of hibernation for the purpose of discovering and characterizing Near Earth Objects (NEOs) and other small bodies. The survey was restarted on December 13th, 2013, and the \neowise{} mission has been ongoing ever since \citep{Mainzer2014}.

\par When \wise{} was in its cryogenic phase, the focal planes operated at temperatures of 30-34 K for the W1 and W2 bands and $\sim7.8$ K for the W3 and W4 bands \citep{Wright2010}. \neowise{} continued operating at the new equilibrium temperature of $\sim 74$ K, allowing the use of the W1 and W2 bands for all future observations. The W2 band spans strong gas emission lines from CO and CO$_2$ and so can be used to detect these species from space. Because W2 spans both CO and CO$_2$ emission lines (denoted as CO+CO$_2$), the two species cannot be differentiated by the \wise{}/\neowise{} photometry alone. Therefore we use \COprod{} as a proxy for the production of both species, since the emission line of CO$_2$ is \replaced{$\sim 12$}{$\sim 11.6$} times stronger than the CO line \citep{1983A&A...126..170C,Bauer2015}. From this point forward the production rate of CO+CO$_2$ (units kg/s) from comets will be referred to as \COprod{}.

\section{Observations}

The \wise{} spacecraft captures images every 11s with the active bands operating in unison and observing the same field of view using a beamsplitter \citep{Wright2010}. The telescope orbits Earth in a pole to pole orbit near terminator and advances on the sky 1$\deg$ per day achieving full sky coverage roughly once every 6 months. This yields an average of $\sim 12$ exposures spaced over $\sim 36$ hours for most comets in the survey \citep{Mainzer2011a,Cutri2012}. The spacecraft may later detect the comet at a different part of its orbit. As in \citet{Bauer2015}, all comet observations with multiple sets of detections within a given date range are referred to as epochs. The separation time between any two epochs is a minimum of $\sim3$ \added{days}. The epochs are numbered in the order of which the comets were observed by the telescope. For example, if there are three different observations of the same comet, each observation will be chronologically labeled epoch 1, 2, and 3. If an observation is not included due to either clear comet inactivity or low SNR, the epoch is skipped. Each band (W1, W2, W3, and W4) has a different spatial resolution, 6.1, 6.4, 6.5, and 12.0 arcseconds, respectively, given as the FWHM of the mean point spread function \citep{Wright2010,Cutri2012}. The images we used were limited to detections of the comet with signal to noise ratios (SNR) $\geqslant 2.5$. All images used in this study were stacked single-exposure images produced by the WISE data pipeline. Of the objects used in this study, 10 have images gathered from the reactivated mission, while 5 objects have images gathered from the prime mission. All comet orbital properties can be found in Table \ref{tab:Properties}, including the breakdown of objects with multiple visits. The date given for each epoch is the median time stamp of the stacked exposures in units of Modified Julian Date (MJD).

\begin{deluxetable*}{lCCCCCC}[ht!]
\tabletypesize{\footnotesize}
\tablecaption{Orbital Properties of HTCs\tablenotemark{a} \label{tab:Properties}}
\tablewidth{0pt}
\tablehead{
Object & \colhead{i} & \colhead{e} & \colhead{q} & \colhead{Phase Ang} & \colhead{Image Stack Mid-point}\\
\colhead{} & \colhead{(deg)} & \colhead{} & \colhead{(AU)} & \colhead{(deg)} & \colhead{(MJD)}
}
\startdata
27P (Crommelin)                     	& 28.97 & 0.9190 & 0.748 & 10.8 & 56719.2097 \\
P/2006 HR30 (Siding Spring)         	& 31.88 & 0.8431 & 1.226 & 6.40 & 55233.4654 \\
P/2010 JC81\tablenotemark{b}        	& 38.69 & 0.7773 & 1.811 & 15.0 & 55327.4473 \\
C/2010 L5 Epoch 1\tablenotemark{b}  	& 147.1 & 0.9037 & 0.791 & 57.5 & 55361.4543 \\
C/2010 L5 Epoch 2\tablenotemark{b}  	& 147.1 & 0.9037 & 0.791 & 39.0 & 55393.9793 \\
P/2012 NJ (La Sagra)\tablenotemark{c}   & 8.503 & 0.8481 & 1.292 & 8.10 & 55311.2952 \\
C/2014 J1 (Catalina) Epoch 2        	& 159.7 & 0.8023 & 1.709 & 36.4 & 56821.8230 \\
C/2014 Q3 (Borisov) Epoch 2         	& 89.95 & 0.9421 & 1.647 & 35.0 & 56936.2784 \\
C/2014 Q3 Epoch 3                   	& 89.95 & 0.9421 & 1.647 & 36.7 & 56989.2271 \\
C/2014 W9 Epoch 2                   	& 10.63 & 0.8578 & 1.587 & 37.6 & 57046.2495 \\
C/2014 W9 Epoch 3                   	& 10.63 & 0.8578 & 1.587 & 38.0 & 57064.9394 \\
C/2015 A1                           	& 80.37 & 0.9008 & 1.996 & 29.8 & 57085.1842 \\
C/2015 GX Epoch 3                   	& 90.25 & 0.8782 & 1.972 & 22.0 & 57100.8549 \\
C/2015 GX Epoch 4                   	& 90.25 & 0.8782 & 1.972 & 26.4 & 57157.8278 \\
C/2015 GX Epoch 5                   	& 90.25 & 0.8782 & 1.972 & 29.4 & 57305.0489 \\
C/2015 H1 (Bressi) Epoch 4          	& 140.7 & 0.9408 & 1.926 & 30.2 & 57156.7226 \\
C/2015 H1 Epoch 5                   	& 140.7 & 0.9408 & 1.926 & 24.9 & 57231.7971 \\
C/2015 X8 Epoch 1                   	& 155.3 & 0.9393 & 1.190 & 43.6 & 57370.8975 \\
C/2015 X8 Epoch 2                   	& 155.3 & 0.9393 & 1.190 & 30.1 & 57422.6712 \\
C/2015 YG1                          	& 57.34 & 0.8792 & 2.073 & 25.9 & 57374.5027 \\
C/2016 S1                           	& 94.69 & 0.7089 & 2.412 & 22.2 & 57723.6497 \\
\enddata
\tablecomments{ \tablenotemark{a} Orbital properties from objects observed during the reactivated phase of the mission except objects from \citep{Bauer2015}. The orbital parameters and phase angles were provided JPL's HORIZONS ephemeris service; \url{https://ssd.jpl.nasa.gov/}. The orbital properties include orbital inclination (i), orbital eccentricity (e), and perihelion distance (q). The Image Stack Mid-Point in MJD was determined by finding the median stacked image date, except for objects provided by \citet{Bauer2015}.\\
\tablenotemark{b} Orbital properties are from \cite{Bauer2015}; these data are objects observed during the prime phase of the mission.
\\
\tablenotemark{c} P/2012 NJ was not included in \citet{Bauer2015}, but was observed during the cryogenic mission.
}
\end{deluxetable*}

\subsection{CO+CO$_2$ Producing HTCs}
When all four bands of \wise{} were operational, reflected light emission was determined from the W1 fluxes, while thermal emission of dust was detectable in the W3 and W4 bands. The W2 ($4.6\pm0.5$ $\mu$m) band is most useful for detecting gas production rates because CO and CO$_2$ emit strong spectral lines at 4.67 $\mu$m and 4.23 $\mu$m respectively \citep{2008AJ....136.1127P,Bauer2011,2013Icar..226..777R}. The spectral line emissions are strong enough to manifest as excess flux in the W2 band, relative to the signal level fit of reflected light and thermal emission present in the W1, W3, and W4 bands, thereby providing a metric for detection. Our sample does not consist of HTCs from the 3-band mission. However, there are 5 objects from the prime mission with measurements made in all four bands (27P, P/2006 HR30, P/2010 JC81, C/2010 L5, P/2012 NJ). To account for observations made by \neowise{}, a method was developed and applied in order to predict the thermal emission curve that was constrained by the signals from the W3 and W4 bands during the prime mission (see Section \ref{Analysis:gas_measure}).

\section{Analysis}

\par In the reactivated \neowise{} Year I and Year II data, certain HTCs were identified as potential candidates for CO+CO$_2$ production by the following method. The images were stacked and photometrically analyzed in the same manner as described in \citet{Bauer2015}. The \wise{} image data were processed using the scan and frame pipeline responsible for applying instrumental, photometric, and astronomical calibrations described in Section IV of \citealt{Cutri2012}. Initial data products were co-added using the software suite known as ``A \wise{} Astronomical Image Co-Adder'' (AWAIC) that takes advantage of advanced interpolation methods in order to maximize SNR \citep{Masci2009}. Emission flux was obtained by converting from signal count using the magnitude zeropoints corresponding to each W-band and 0th magnitude flux values procured from \citet{Wright2010}. Color corrections and aperture photometry were performed on the stacked images with an 11 arcsecond radius.  To assure that a stacked image contained a comet, and not a star or noise, we completed by-eye visual inspections of the selected HTCs with SNRs $\geqslant$ $\sim2.5$. At the completion of visual assessment, it was found that we had 11 candidates from the reactivated mission. Images of the HTCs 27P, P/2006 HR30, and P/2012 NJ from the prime mission were also used in this study for completeness. In addition, two epochs of C/2010 L5, and a single epoch of P/2010 JC81 were included from \citet{Bauer2015}.

\subsection{Nucleus Size}

\par The methods used to determine nucleus size were similar to those described for comets in \cite{Bauer2015}, and those described for asteroids in \cite{Masiero2012}, \cite{2015ApJ...814..117N}, and \cite{Nugent_2016AJ}. The diameter of the comets 27P, P/2006 HR30, P/2012 NJ, and C/2016 S1 were calculated using aspects of the Near-Earth Asteroid Thermal Model (NEATM) first described in \cite{1998Icar..131..291H}. The model assumes a spherical object, with no rotation, no night side emission, and a temperature distribution given by Equation~\ref{Eq:temp_dist}.

\begin{equation}\label{Eq:temp_dist}
T(\theta)=T_{max}\cos^{1/4}(\theta)\textrm{ for }0\leqslant \theta \leqslant \pi/2
\end{equation}

For the above equation, $\theta$ is the angular distance from the sub-solar point and $T_{max}$ can be defined as the sub-solar temperature given by Equation~\ref{Eq:t_max}, in which $S$ is defined as the solar flux at the asteroid, $\eta$ is the beaming parameter as described in \citet{1998Icar..131..291H}, $A$ is the bolometric Bond albedo, $\epsilon$ is the emissivity, and $\sigma$ is the Stefan-Boltzmann constant.

\begin{deluxetable*}{lCCCCCCCCC}[ht!]
\tabletypesize{\footnotesize}
\tablecaption{Dust and $W2$ Excess Analysis Results \label{tab:Production}}
\tablewidth{0pt}
\tablehead{
Object & \colhead{R$_\textrm{h}$} & \colhead{Delta} & \colhead{$Q$\textsubscript{CO\textsubscript{2}}} & \colhead{$Af\rho$} & \colhead{CO2/$Af\rho$} & \colhead{$Q$\textsubscript{CO\textsubscript{2}} $3\sigma$ Upper}\\
\colhead{} & \colhead{AU} & \colhead{AU} & \colhead{molecules/cm} & \colhead{cm} & \colhead{molecules/cm$\cdot$s} & \colhead{molecules/s}}

\startdata
27P (Crommelin)                    & 5.39 & 5.31 &               &              &              & 26.3 \\
P/2006 HR30 (Siding Spring)        & 8.81 & 8.77 &               &              &              & 26.8 \\
P/2010 JC81						   & 3.90 & 3.76 &				 &				&			   & 25.5 \\
C/2010 L5 Epoch 1                  & 1.21 & 0.65 & 26.71\pm0.25  & 1.95\pm0.010 & 24.38\pm0.25 &      \\
C/2010 L5 Epoch 2.                 & 1.62  & 1.15 & 25.08\pm0.08  & 2.64\pm0.14  & 22.44\pm0.16 &      \\
P/2012 NJ (La Sagra)               & 7.08 & 6.94 &               &              &              & 26.6 \\
C/2014 J1 (Catalina) Epoch 2       & 1.71 & 1.35 & 25.18\pm0.12  & 0.790\pm0.036& 24.39\pm0.13 &      \\
C/2014 Q3 (Borisov) Epoch 2        & 1.74 & 1.40 & 26.71\pm0.12  & 2.57\pm0.010 & 24.13\pm0.12 &      \\
C/2014 Q3 Epoch 3                  & 1.65 & 1.32 &               & 2.56\pm0.010 &              & 26.8 \\
C/2014 W9 Epoch 2                  & 1.61 & 1.26 &               & 1.57\pm0.050 &              & 26.1 \\
C/2014 W9 Epoch 3                  & 1.59 & 1.12 & 26.16\pm0.11  & 1.69\pm0.011 & 24.47\pm0.11 &      \\
C/2015 A1                          & 2.00 & 1.73 &               & 1.51\pm0.010 &              & 25.8 \\
C/2015 GX Epoch 3                  & 2.66 & 2.46 & 25.54\pm0.12  & 1.31\pm0.090 & 24.24\pm0.15 &      \\
C/2015 GX Epoch 4                  & 2.29 & 2.05 & 25.81\pm0.10  & 1.30\pm0.046 & 24.51\pm0.11 &      \\
C/2015 GX Epoch 5                  & 2.04 & 1.77 & 26.40\pm0.10  & 1.98\pm0.096 & 24.41\pm0.14 &      \\
C/2015 H1 (Bressi) Epoch 4         & 2.01 & 1.72 & 25.83\pm0.12  & 1.73\pm0.031 & 24.10\pm0.12 &      \\
C/2015 H1 Epoch 5                  & 2.40 & 2.09 & 25.58\pm0.12  & 1.59\pm0.040 & 23.99\pm0.13 &      \\
C/2015 X8 Epoch 1                  & 1.43 & 1.03 & 26.00\pm0.10  & 1.81\pm0.056 & 24.19\pm0.11 &      \\
C/2015 X8 Epoch 2                  & 1.92 & 1.47 & 25.76\pm0.12  & 1.56\pm0.057 & 24.20\pm0.13 &      \\
C/2015 YG1                         & 2.26 & 2.04 & 26.24\pm0.12  & 1.86\pm0.15  & 24.38\pm0.19 &      \\
C/2016 S1                          & 2.60 & 2.34 &               &              &              & 25.5 \\
\enddata
\tablecomments{R$_\textrm{h}$ is the comet-sun distance; Delta is the comet-Earth distance.}
\end{deluxetable*}

\begin{equation}\label{Eq:t_max}
T_{max}=\left(\frac{(1-A)S}{\eta \epsilon \sigma}\right)^{1/4}
\end{equation}

\par NEATM was applied to our comets under the assumption that the coma was inactive and the object appeared point-like. We made this assumption due to the surface brightness profile of the comet matching that of a point spread function for \wise{}. For any given object we only have thermal data from the W2 band, thus any aberration between NEATM and the actual comet is accounted for using the beaming parameter $\eta$ \citep{Masiero2012}. Such aberrations comprise of non-spherical shapes, variations in surface roughness or thermal inertia, the presence of satellites, uncertainties in emissivity, high rates of spin, changes in surface temperature distributions due to spin pole location, or the imprecise assumption that the object’s night-side has zero thermal emission \citep{2015ApJ...814..117N}; for example $\eta = \pi$ is representative of a spherical body with high thermal inertia \citep{1998Icar..131..291H}. When generating the thermal profile, a crucial step in determining the albedo and diameter of the comet, $\eta$ can be used to modify the temperature distribution of the model to account for any of these anomalies. For our analysis the beaming parameters were fixed if only one detection band had an acceptable SNR or if there was a detection in only the W1 band or the W2 band. In every other case $\eta$ was left a free parameter and ranged from 0.63 to 1.3.

\subsection{Dust Photometry and CO+CO$_2$ Production Measurements} \label{Analysis:gas_measure}

The quantity $Af\rho$, as determined in \citet{Bauer2015} and defined in \citet{AHearn1984}, was calculated as a benchmark to compare comets within the HTC population. The quantity is calculated from the reflected light component of emission, presumably from the dust coma of the comets. $Af\rho$ consists of the albedo $A$, the filling factor $f$, and the aperture radius projected out to the distance of the comet $\rho$ \citep{AHearn1984}. The filling factor is $f=\textrm{\Ndust} \sigma/\pi\rho^{2}$ with \Ndust being the number of dust grains in a given aperture and $\sigma$ being the grain cross section.

\par $Af\rho$ can also be defined in the form of Equation~\ref{afrho}, in which $F_{\textrm{comet}}$ is the observed flux from the comet, $F_{\astrosun}$ is the flux from the Sun evaluated at 1 AU, $r$ is the Sun-comet distance in AU, and $\Delta$ is the Earth-comet distance in the same units as $\rho$ (for ground-based observations), or in this case the distance between the comet and \neowise{} in the same units as $\rho$. The flux from the comet per unit area can also be thought of as $F/4\pi \Delta^2$, with $F$ being the total cometary flux. The data provided by the W1 band were used to calculate $Af\rho$. $Af\rho$ is not independent of phase angle, and is known to affect dust production calculations \citep{ahearn1995}. To account for phase angle, a correction was made using the phase function as defined by \citet{Agarwal2007}. The computation of $Af\rho$ is a necessary step in calculating \COprod.

\begin{equation} \label{afrho}
	Af\rho=\frac{F_{\textrm{comet}}}{F_{\textrm{\astrosun}}}\frac{(2r\Delta)^2}{\rho}
\end{equation}

\par \deleted{The computation of $Af\rho$ is a necessary step in calculating \COprod.}The production rate of CO+CO$_2$ was calculated for objects observed during the prime mission from the excess signal $(\sim3 \sigma)$ in the W2 band relative to the extrapolated thermal and reflected light contributions. The thermal contribution was determined by fitting a Planck function for an appropriate temperature to the W3 and W4 fluxes. Detections for our objects in the W3 and W4 bands were absent for the post-cryo and reactivation mission phases. \deleted{thus we had to estimate the curve for thermal emission} \added{Thus, the theoretical thermal emission curve was determined by calculating the expected blackbody radiation emission given the estimated amount of dust from the W1 flux.} Assuming that dust grains both reflect light and emit thermally, we used the number of dust grains \replaced{$N$}{\Ndust} to find a number surface density. Given the number density, a Planck function was generated using an infrared emissivity of $\sim 0.9$, assuming a dust grain albedo near $\sim0.1$ and assuming a blackbody temperature of $286\textrm{K} \times r_H^{-1/2}$ in which $r_H$ (AU) is the heliocentric distance \citep{Stevenson2015}.

\par
\begin{figure}[H]
\hspace*{-1.95cm}
\includegraphics[scale=0.4]{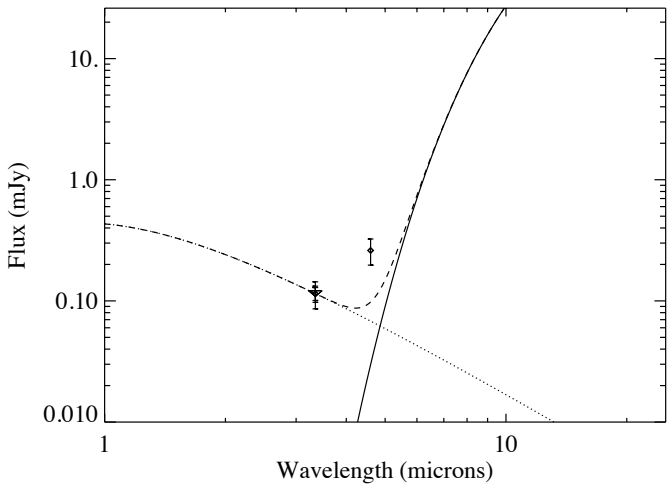}
\caption{\footnotesize{Example of 4.6 $\mu$m excess in the thermal signal for C/2015 H1 Epoch 2. The reflected light emission flux as a function of wavelength is represented by the dotted line and is constrained by data in the W1 band. The thermal emission flux as a function of wavelength is represented by the solid line and is approximated from data in the W1 band. The combined reflected light emission and thermal emission curve is represented by the dashed line. Note the 4.6 $\mu$m excess in combined reflected and thermal emission flux detected in the W2 band ($\blacklozenge$). Note that flux distributions are not shown for the comets 27P, P/2006 HR30, P/2012 NJ, and C/2016 S1, which appeared to be inactive, i.e did not exhibit detectable coma, during the time of the NEOWISE observations. Flux distributions for C/2010 L5 can be found in \citealt{Kramer2017}.} The complete set of plots is available in the online journal.}
\label{DetectPlot}
\end{figure}

 For each object, a plot of the detected reflected light flux (assumed to be a solar spectrum shape) and the estimated thermal signal was produced. The reflected light curve was constrained by detections in the W1 band. If there was excess flux detected by W2, such as in Figure \ref{DetectPlot}, we assumed that this was due to \COprod. This simplification was necessary due to the single data point from the reflected emission at 3.4 $\mu$m and due to the absence of longer wavelength data. \added{The excess W2 signal flux, is converted to an average column density $\braket{N}$ in units of cm$^{-2}$ by way of Equation~\ref{column_dens}. 

\begin{equation}\label{column_dens}
  \braket{N} = F_{\textrm{\scriptsize W}_2}4\pi \Delta^2 \frac{\lambda}{hc}\frac{r^2}{g}\frac{1}{\pi\rho^2}
\end{equation}

\par The variables present from Equation~\ref{afrho} share the same definition, $F_{\textrm{\scriptsize W}_2}$ is the total W2 excess flux in units of erg s$^{-1}$ cm$^2$, $\lambda$ is the wavelength of the observation in units of $\mu$m, $h$ is the Planck constant in units of erg s, $c$ is the speed of light in units of $\mu$m, and $g=2.86\times10^{-3}$ s$^{-1}$ is the fluorescence efficiency for CO$_2$ at 1 AU \citep{1983A&A...126..170C}. We calculate $F_{\textrm{\scriptsize W}_2}$ after eliminating the inband nucleus dust signal contributions and integrating the resultant flux density over the CO/CO$_2$ bandpass. 

\par Knowing $\braket{N}$ we can calculate \COprod{} by Equation~\ref{co_prod}, where $v$ is the gas ejection velocity assumed to be 0.62 km s$^{-1}$ consistent with \citet{Bauer2011}, and $10^5$ is a unit conversion factor.} In order to calculate uncertainties for \COprod we accounted for both the uncertainty from the photometry, and the uncertainty in the thermal emission calculation; both were added in quadrature.
\added{
\begin{equation}\label{co_prod}
	Q_{\textrm{\scriptsize CO}_{2}} = \braket{N}2\rho v \times 10^5
\end{equation}}

\par In some cases only an upper limit of \COprod could be calculated for the objects. We found this value by assuming that the comet was active and dominated by the dust emission signal. Furthermore, we found the value $1 \sigma$ above and below the W2 band signal and considered the amount of CO$_2$ that would be produced given the blackbody estimate and given the scattered reflected light across the solar spectrum. If the $1\sigma$ photometric uncertainty fell on the curve created from the sum of the thermal emission curve and reflected light curve at W2, we calculated an upper limit. The combined photometry of the upper $1 \sigma$ uncertainty and the W2 band signal was then used to find the $1\sigma$ \COprod upper limit. For the purposes of our study, the $1 \sigma$ upper limits were converted to $3 \sigma$ upper limits.

\par When calculating either \COprod{} or \COprod{} upper limits we convert all excess flux into equivalent CO$_2$ production rates. However, if the excess is attributed solely to the production of CO, the equivalent CO production rates can be obtained by multiplying \COprod{} by a factor of \replaced{12}{$\sim11.6$}. Note that if CO is the dominant source of the emission, a mix of the two is very possible (cf. \citealt{2016MNRAS.462S.156F}). \replaced{The overall process used to calculate $Af\rho$ and \COprod was very similar to that developed by \citet{Bauer2015}.}{Overall, the process to calculate \COprod builds upon methods introduced by \citet{2008AJ....136.1127P} and is very similar to that developed in \citet{Bauer2011, Bauer2012a, Bauer2012b, Bauer2015, Stevenson2015}.}


\par C/2010 L5 Epoch 2 was not considered for this analysis due to its low $Af\rho$ value and further considerations from \citep{Kramer2017} that indicate the comet underwent a single pathological outburst event without continuing activity in Epoch 2.

\par It was found that significant \COprod{} signal ($3\sigma$) was found in all 11 of the reactivated mission's HTCs, shown in Table \ref{tab:Production}. We have a total of 13 measurements of \COprod{} when including the two epochs of C/2010 L5 \citep{Bauer2015}. There were 8 other HTCs that demonstrated limited to no activity. However, we were able to calculate $3\sigma$ upper limits for \COprod for each comet because the surface brightness profile matched the stellar point-spread function within the limits of uncertainty.

\section{Discussion}

\subsection{Nucleus Size}

\par The comet diameters in Table~\ref{tbl:nucleus} provide a representation of the sizes of comets used in this study. The range in diameters is similar to those of long period comets. Due to concerns with the degree of activity of C/2016 S1 we calculated a $3\sigma$ upper limit diameter of 14 km. However, if our observation of C/2016 S1 was during a period of inactivity, we derived the diameter of the comet to be 5.2$\pm$3 km. C/2010 L5 also has a calculated $3\sigma$ upper limit of 2.2 km \citep{Kramer2017}. It should be noted that the nucleus size of C/2010 L5 is much smaller than that of the other HTCs. In general, the nucleus mean diameter of the HTCs in this sample are on the order of Halley's effective diameter of 11 km \citep{cometsII}.

\begin{deluxetable}{lCC}[h]
\tabletypesize{\footnotesize}
\tablecaption{HTC Nuclei Diameters\tablenotemark{a} \label{tab:Production}}
\tablewidth{0pt}
\tablehead{
Object & \colhead{Diam (km)} & \colhead{$\eta$}}

\startdata
27P                         & 12\pm3    & 1.0, fixed \\
P/2006 HR30                 & 16\pm2    & 0.63\pm0.1 \\
C/2010 L5\tablenotemark{a}  & \leq2.2   &  $-$       \\
P/2012 NJ                   & 19\pm2    & 1.3\pm0.2  \\
C/2016 S1                   & 5.2$\pm$3 & 1.0, fixed \\
\enddata
\tablecomments{
\tablenotemark{a} $\eta$ is the beaming parameter.\\
\tablenotemark{b} Diameter from \citet{Kramer2017}.}
\label{tbl:nucleus}
\end{deluxetable}

\subsection{Dust Production Rates and CO+CO$_2$ Production Rates}\label{Disc:CO2}

\par In order to examine the relationship between our independent variables ($Af\rho$, \COprod{}, \COprod{}/$Af\rho$) and $R_h$, we calculated both the Spearman rank correlation coefficient ($\rho_s$) and the Kendall-$\tau$ ($\tau$). The strength and direction of the monotonic relationship between two continuous or ordinal variables is described by $\rho_s$, and can range from $-1$ to $1$. A value of $-1$ indicates a perfect negative correlation and $1$ indicates a perfect positive correlation, a value of $0$ indicates no correlation between variables. The significance of $\rho_s$ was found by calculating the two-tailed $p-value$. A $p-value$ for $\rho_s$ can range from 0 to 1, the closer the value is to 0, the greater the significance of the result. It should be noted that the $p-value$ calculated for a given $\rho_s$ may not be reliable for a sample of our size, thus a $\tau$ test was also implemented to our data. The main difference between $\tau$ and $\rho_s$ is that the calculations for $\tau$ are based on concordant and discordant pairs in the data, and the calculations for $\rho_s$ are based on deviations in the data. This fact makes the $\tau$ test more accurate for smaller sample sizes. Like $\rho_s$, $\tau$ can range from -1 to 1; a value of -1 indicating a negatively correlated pair, 0 indicating no correlation, and 1 indicating a positively correlated pair. Again, the significance of the test was determined by calculating a two-tailed $p-value$. Originally, polynomial fits of the data were attempted; however, our sample size was too small and our error too large to meaningfully constrain or identify any power law relationship.

\par
\begin{figure}[h]
\includegraphics[width=0.5\textwidth]{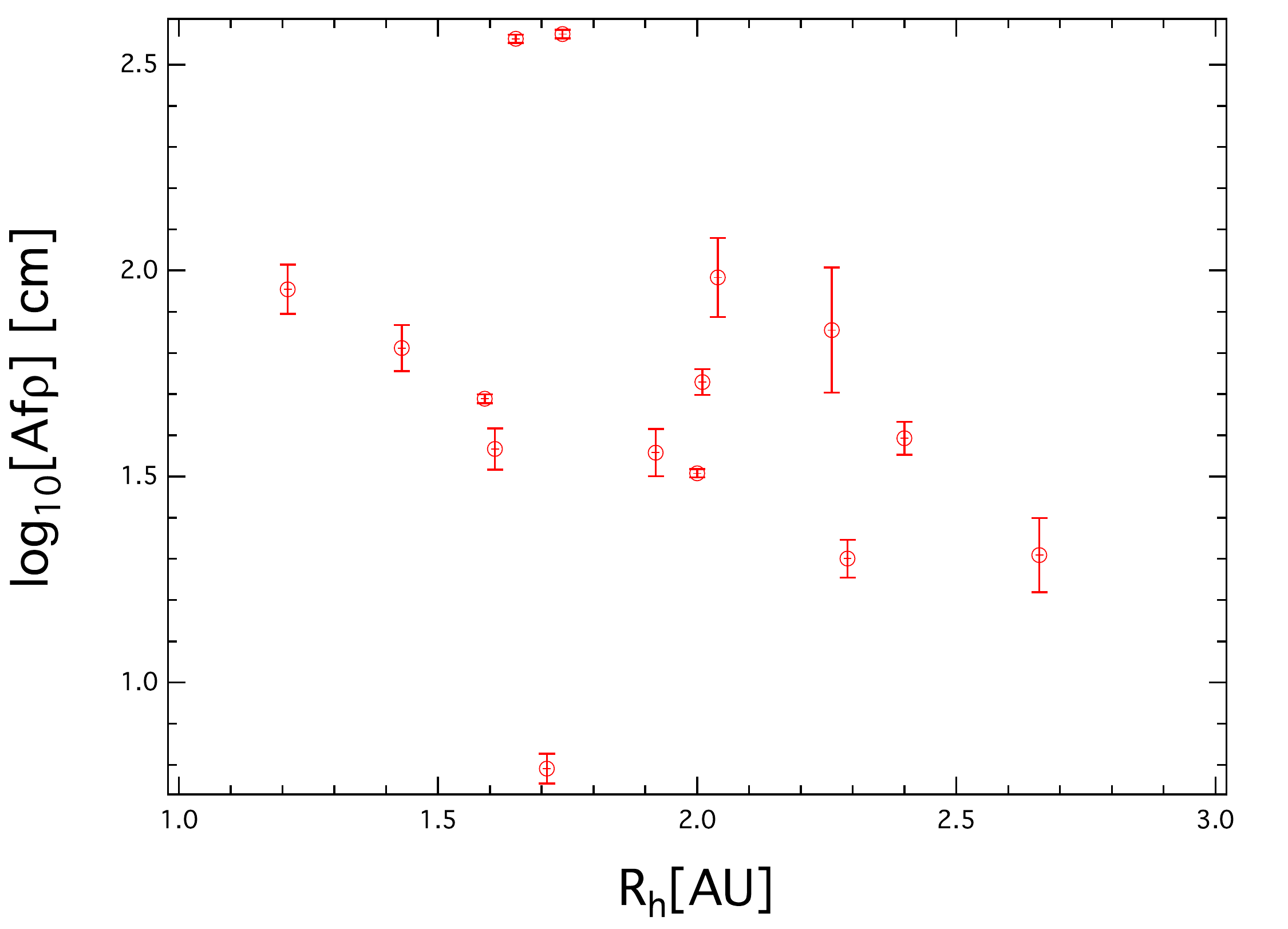}
\caption{\footnotesize{Plot of $Af\rho$ with respect to $R_{Helio}$ derived from 3.4 $\mu$m flux for active HTCs.}}
\label{Fig:afrho}
\end{figure}

 The relationship between $R_h$ from 1.21 to 2.66 AU and dust production rates of the HTCs is shown in Figure~\ref{Fig:afrho}. Comets with a lower $Af\rho$ can be thought of as relatively dust poor or inactive. The rank correlation tests returned a value of $\rho_s=-0.32$ with a $p-value=0.25$, and a value of $\tau=-0.24$ with a $p-value=0.22$, indicating that there is little to no significant correlation between $Af\rho$ and $R_h$ within 1.21 and 2.66 AU.

\par
\begin{figure}[htb!]
\includegraphics[width=0.5\textwidth]{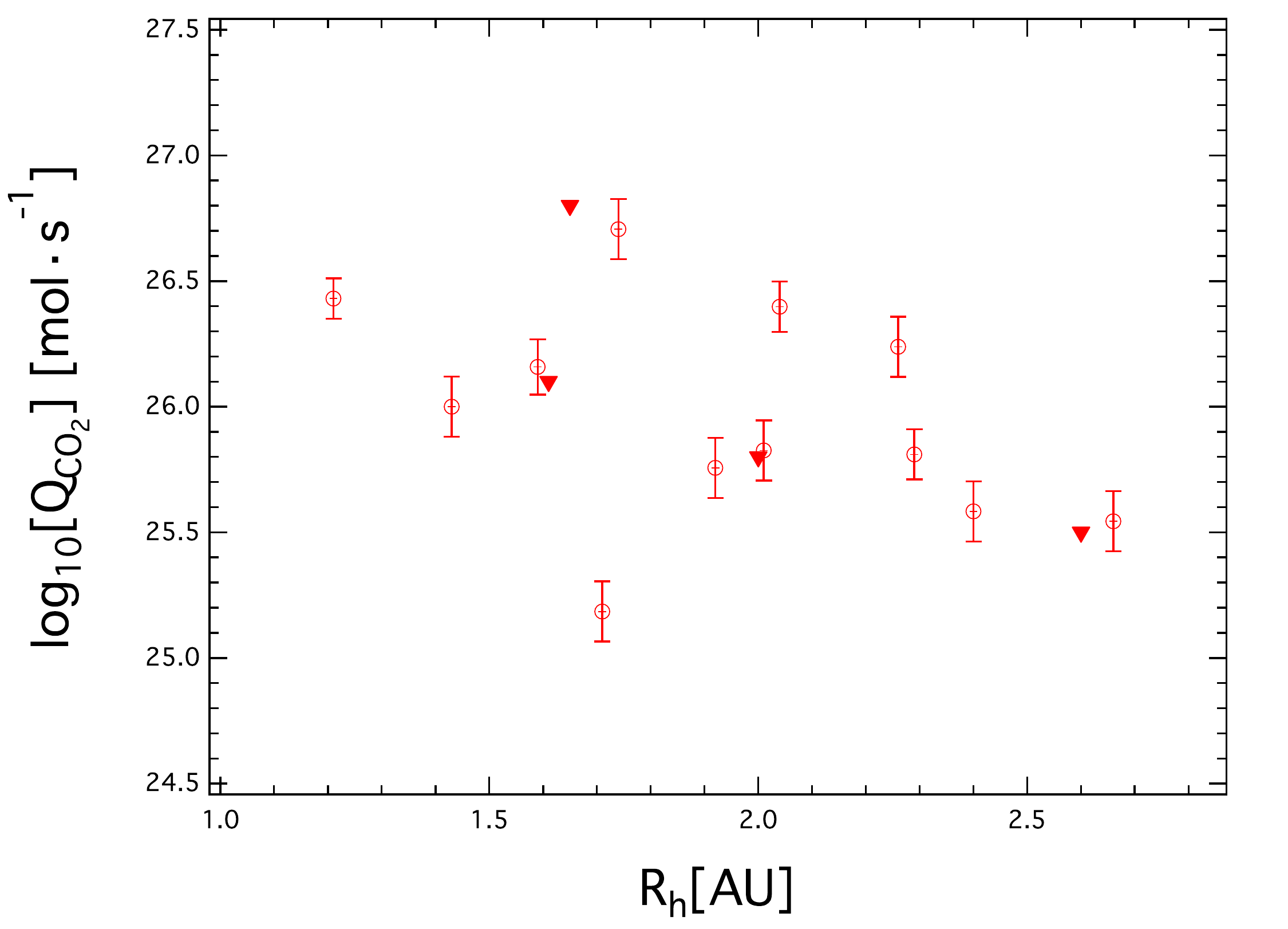}
\caption{\footnotesize{Plot of \COprod{} of HTCs (red circles) with respect to heliocentric distance with Q$_\textrm{CO2}$ upper limits ($\blacktriangledown$) included.}}
\label{Fig:CO2}
\end{figure}

 The relationship between \COprod{} and $R_h$ can be found in Figure \ref{Fig:CO2}. The statistics only consider known \COprod{} values and not the calculated $3\sigma$ upper limits. The rank correlation tests for \COprod{} returned a value of $\rho_s=-0.42$ with a $p-value=0.17$ and a value of $\tau=-0.33$ with a $p-value=0.13$.

\noindent Again, this indicates that there is little to no significant correlation between \COprod{} and $R_h$ for the sample range. For the upper limit of CO$_2$ production rates for C/2014 Q3, C/2014 W9, and C/2015 A1 it should be noted that $R_h$ was at or under 2 AU. For this reason, it is clear from the flux plots in the Appendix that the W2 flux is ever so slightly higher than the thermal curve, but still within the limits of the combined $3\sigma$ uncertainty of the photometry and the model. Due to the proximity to the Sun, the W2 band may have a more significant thermal signal relative to the reflected light signal, leading to a non-optimal prediction of the Planck function. As previously mentioned, the thermal curves are predictions modeled with assumptions from the W1 band data because of the unavailability of the W3 and W4 band data. For this reason there could be a more significant thermal component than the assumed model would suggest. Due to the relatively large heliocentric distances of 27P, P/2006 HR30, C/2010 JC81, and P/2012 NJ, the respective upper limits were not significant and fell outside the plotted range of $R_h$ in Figure \ref{Fig:CO2}. At these distances the activity of the comet could be considered minimal.

\begin{figure}[htb!]
\includegraphics[width=0.5\textwidth]{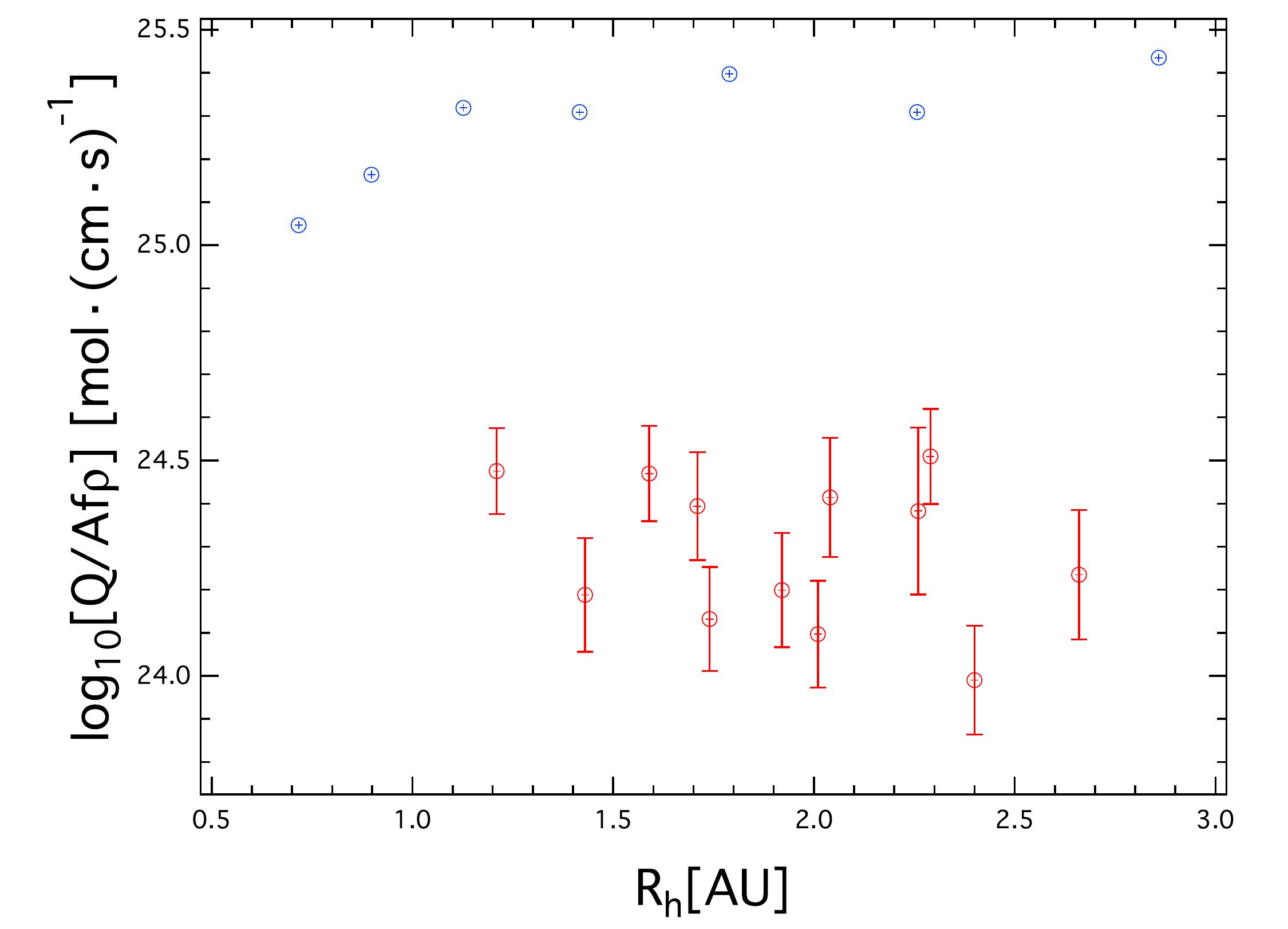}
\caption{\footnotesize{Plot of \coafrho (red circles) as a function of the heliocentric distance. The points in blue are $Q_\textrm{OH}/Af\rho$ with respect to heliocentric distance of 1P Halley as presented by \cite{ahearn1995}.}}
\label{Fig:CO2/afrho}
\end{figure}

\par To consider the nature of dust production and \COprod{} with respect to $R_h$, we plotted the ratio of \COprod{} to $Af\rho$ in Figure \ref{Fig:CO2/afrho}. Our aspiration was to observe if this value indicated any consistency across the HTCs for gas to dust ratios. Again, we note that after completing the rank correlations for \COprod{}$/Af\rho$, we find $\rho_s=-0.20$ with a $p-value=0.53$, and $\tau=-0.18$ with a $p-value=0.41$, indicating that there is little to no correlation with heliocentric distance, but a rather consistent nature of gas to dust production across the objects. It is important to note that, when comparing the behavior of $Q_\textrm{OH}/Af\rho$ for comet 1P/Halley from \citealt{ahearn1995}, the slope of the line is relatively flat for the $R_h$ range of our dataset. We can compare $Q_\textrm{OH}/Af\rho$ for comet Halley with \COprod{}/$Af\rho$ for our comets to find that \COprod{} is roughly $\sim10\%$ on average of $Q_\textrm{OH}$ for the same range in $R_h$. At these distances, then, it is perhaps not surprising that, as with other comets, H$_2$O, the parent of OH, likely dominates activity. For all three cases, the lack of correlation to $R_h$ would indicate that HTCs with  1.21 $\leq R_h \leq 2.66$ AU have production values for both dust and CO+CO$_2$ that are independent of heliocentric distance.


\section{Conclusions}

\par The 11 HTCs analyzed in this study indicate the following:
\begin{enumerate}
	\item We find no significant correlation between heliocentric distance of 1.21 to 2.66 AU and $Af\rho$.
	\item We also find no significant correlation between heliocentric distance of 1.21 to 2.66 AU and \COprod{}.
	\item We also find no significant correlation between heliocentric distance of 1.21 to 2.66 AU and \COprod{}$/Af\rho$, consistent with $Q_\textrm{OH}/Af\rho$ for comet 1P, possibly implying that the rate of dust produced relative to the rate of CO+CO$_2$ produced is independent of the HTC's distance from the Sun.
	\item The behavior observed for this sample of HTCs is not dissimilar from that of the LPCs observed in \citealt{Bauer2015} within 4 AU. However, the results from \citealt{Bauer2015} were primarily for $\epsilon f\rho$ which is an equivalent of $Af\rho$; the difference being $\epsilon f\rho$ is derived from reflected light emission rather than thermal emission.
	\item The newly-derived diameters of HTC's nuclei shown in Table \ref{tbl:nucleus} are in the range of the $\sim 11$ km size found for 1P's nucleus from spacecraft encounters.
\end{enumerate}


\section{Acknowledgments}
This research makes use of the data products from the \textit{Wide-field Infrared Survey Explorer}, a joint venture between the University of California, Los Angeles, and the Jet Propulsion Laboratory/California Institute of Technology, funded by the National Aeronautics and Aerospace Administration. This paper also makes use of data products from \neowise{} a JPL/Caltech project, funded by the Planetary Science Division of NASA.
\added{\software{Astropy \citep{2013A&A...558A..33A}, AWAIC \citep{Masci2009}, IDL (\href{http://www.harrisgeospatial.com/SoftwareTechnology/IDL.aspx}{Exelis Visual Information Solutions}, Boulder, Colorado), Igor Pro (\href{https://www.wavemetrics.com}{WaveMetrics}), Python (\href{https://www.python.org/}{Python Software Foundation}), Pandas \citep{mckinney-proc-scipy-2010}, Scipy \citep{scipy,4160250,5725235}}}
\clearpage


\appendix
\begin{figure*}[ht!]
\gridline{\fig{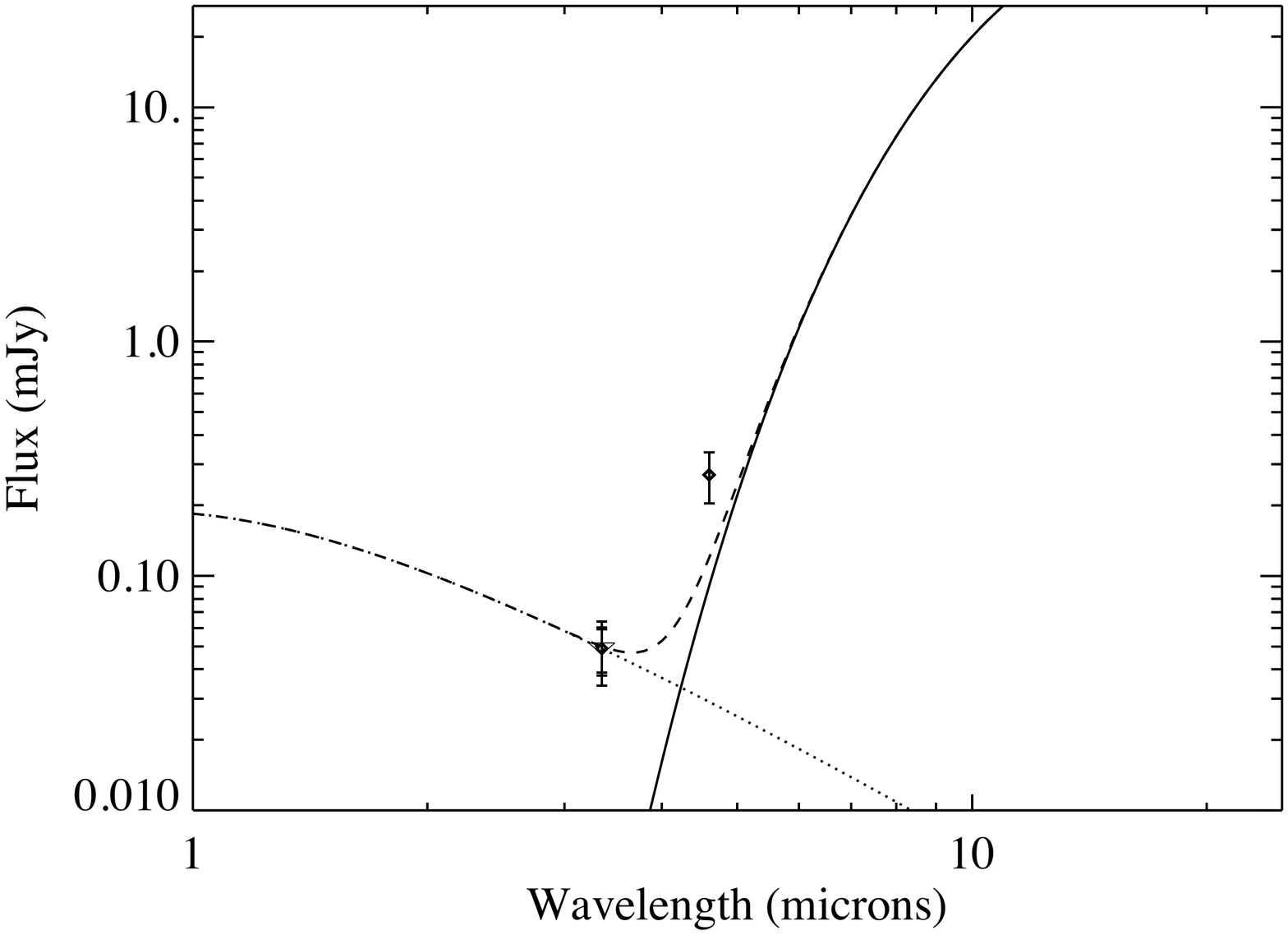}{0.44\textwidth}{(c)}
          \fig{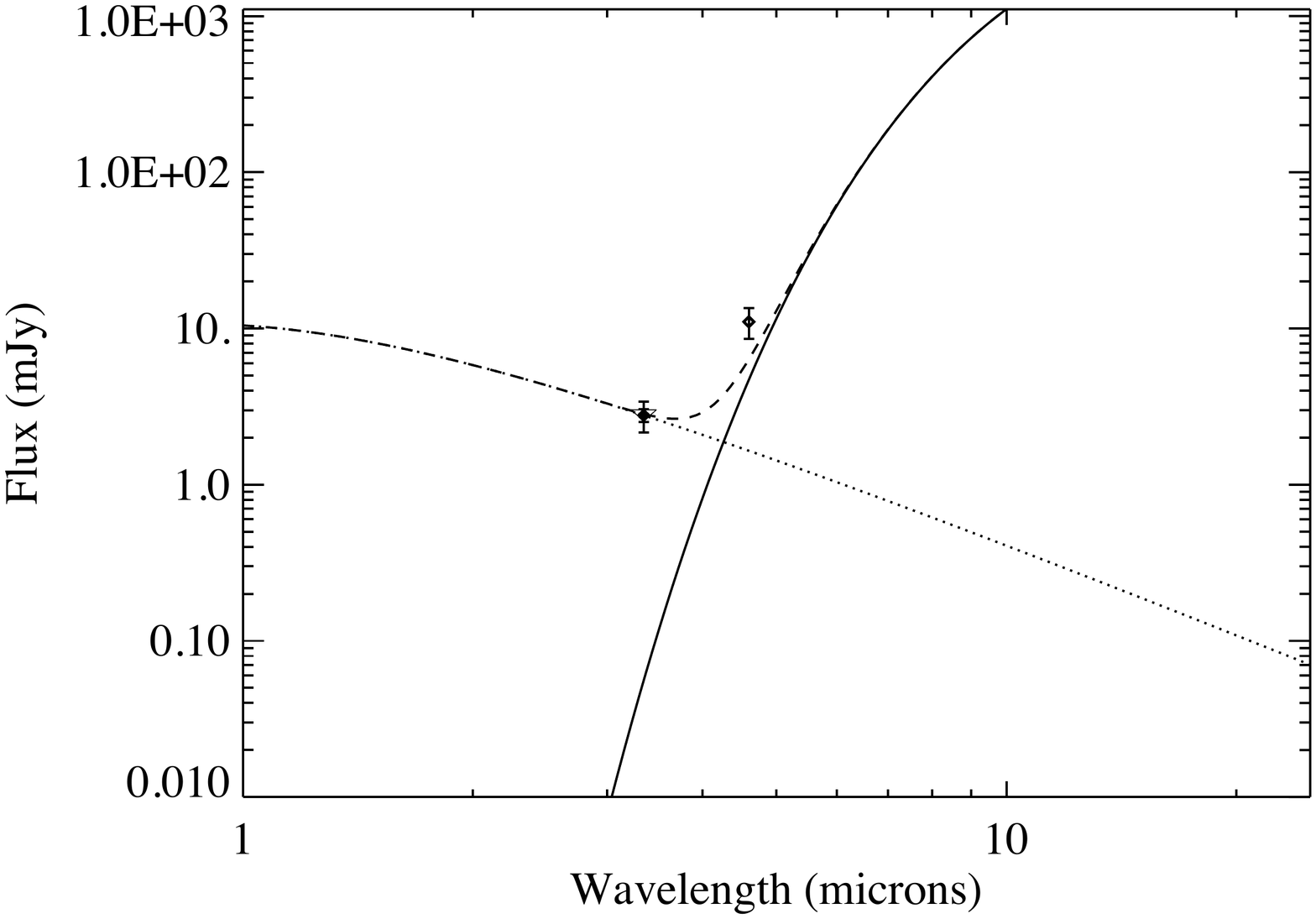}{0.44\textwidth}{(d)}
          }
\gridline{\fig{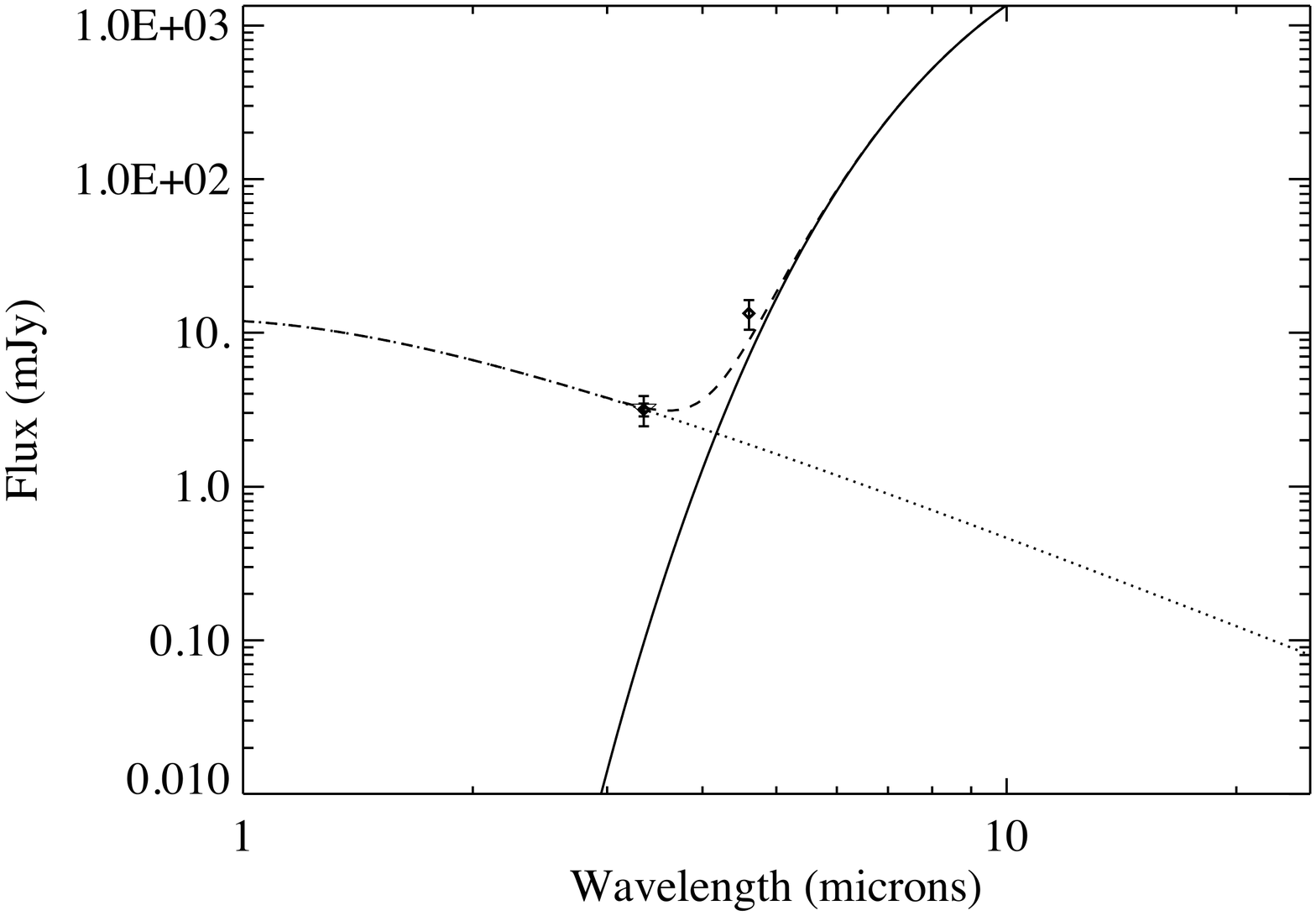}{0.44\textwidth}{(e)}
          \fig{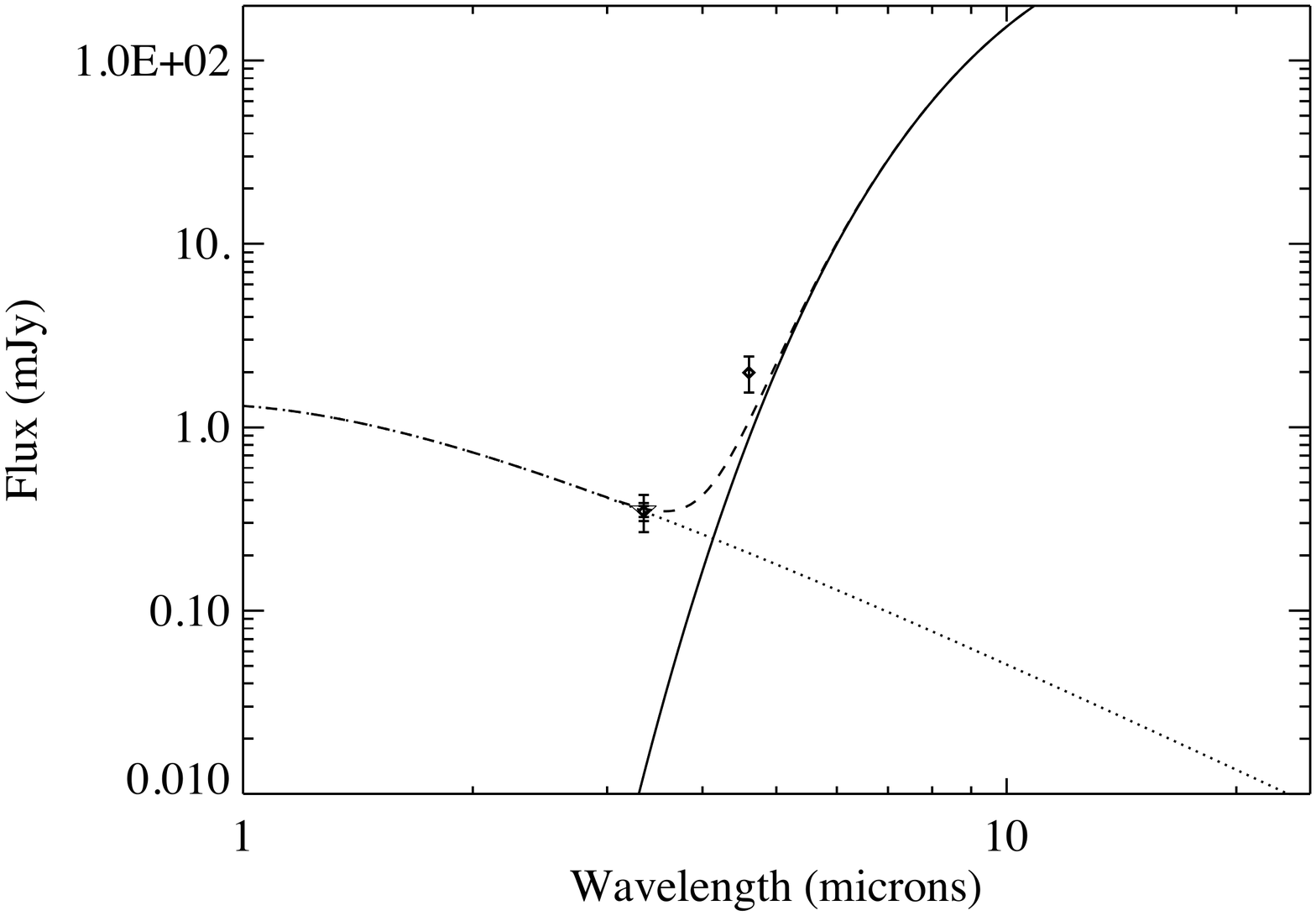}{0.44\textwidth}{(f)}
          }
\gridline{\fig{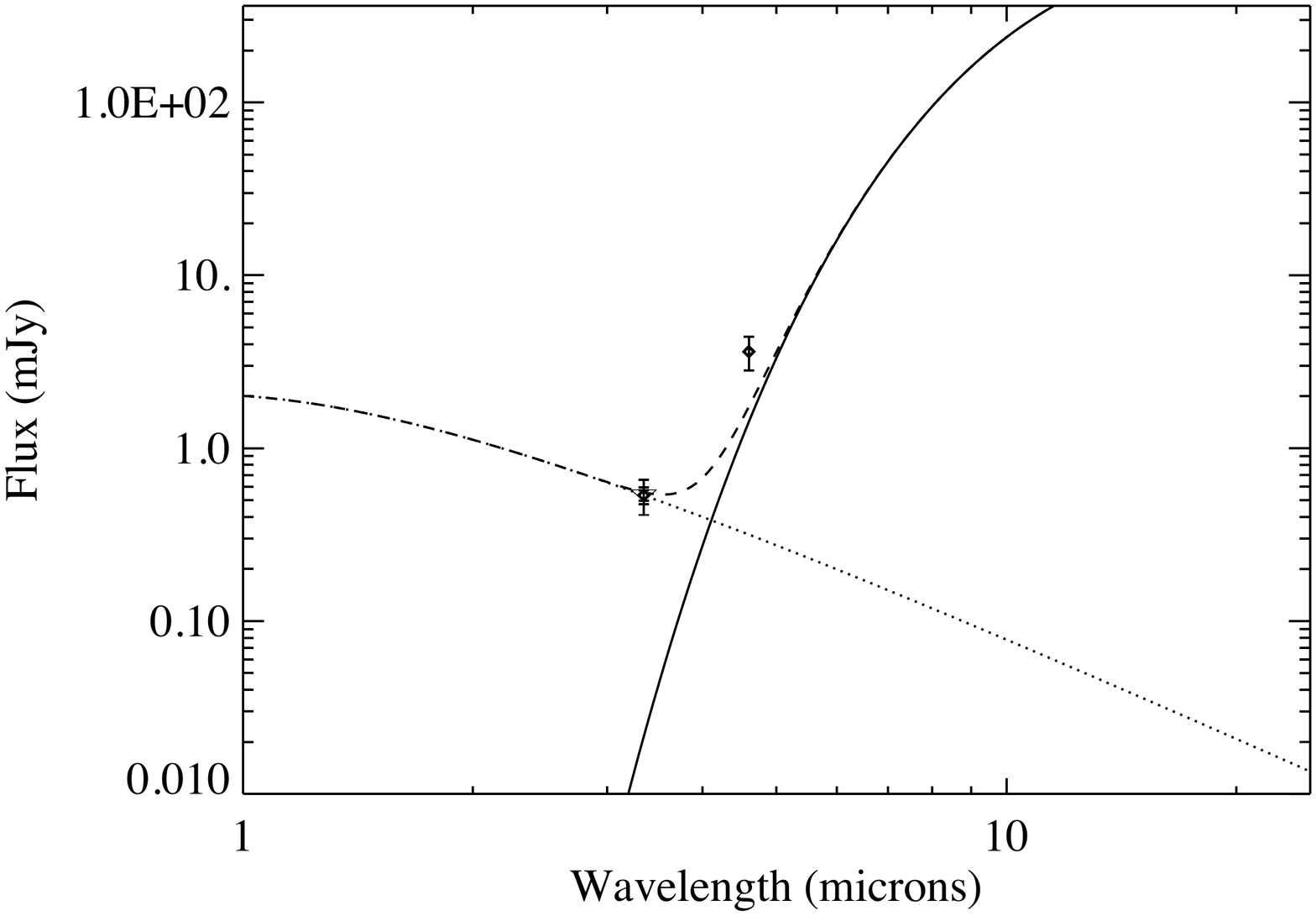}{0.44\textwidth}{(g)}
          \fig{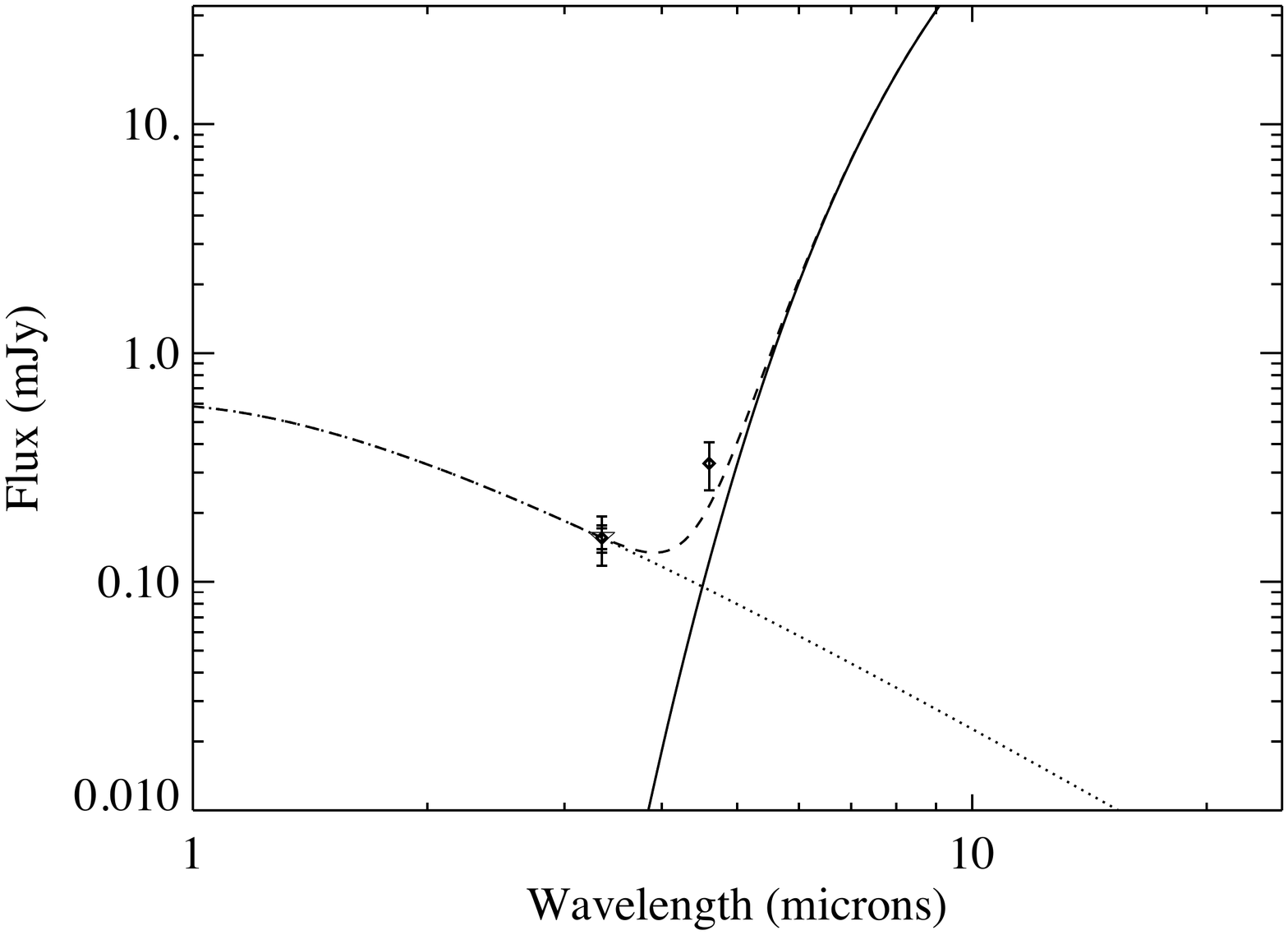}{0.44\textwidth}{(h)}
          }
\end{figure*}

\begin{figure*}[h]

\gridline{\fig{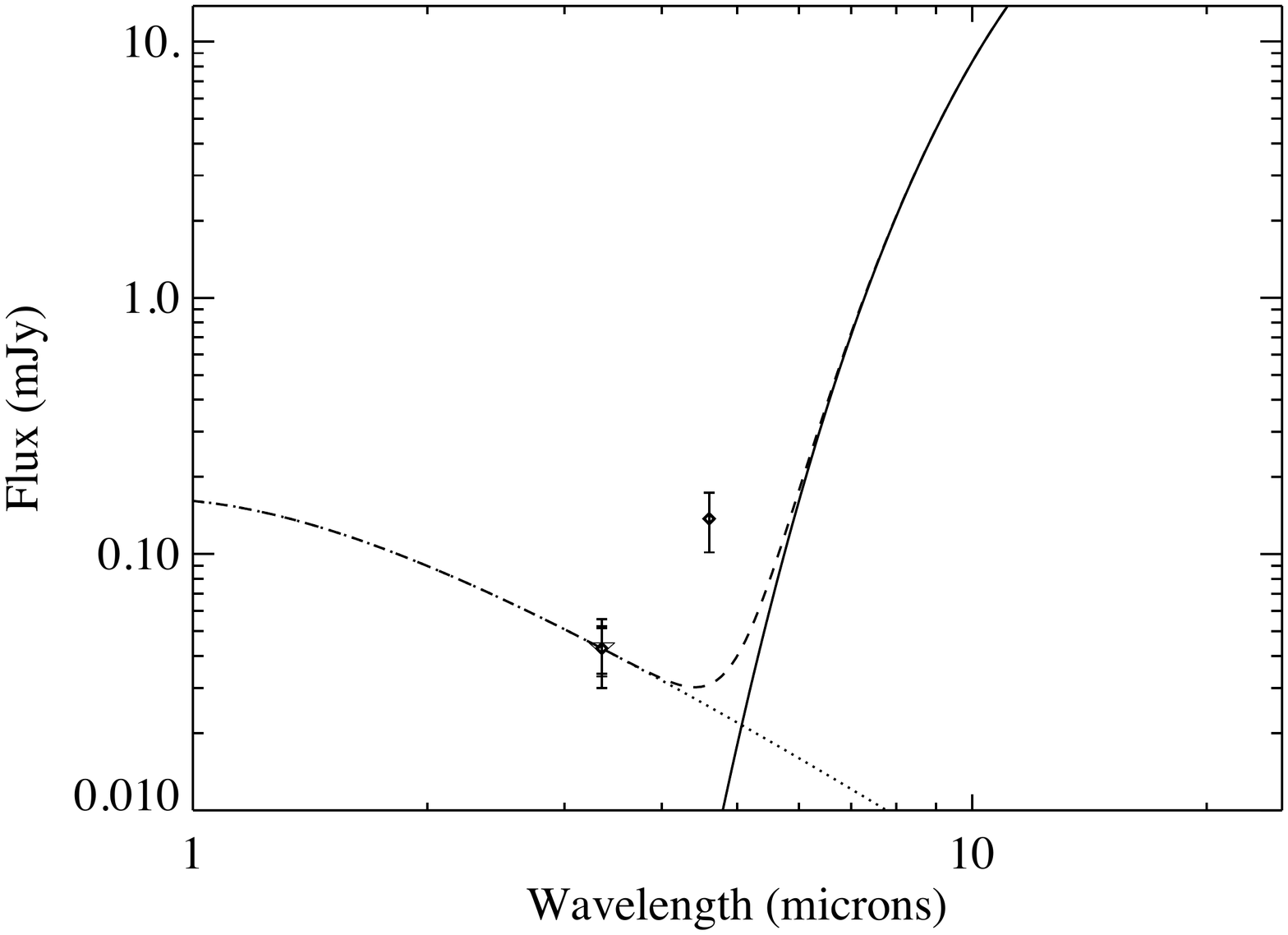}{0.44\textwidth}{(i)}
          \fig{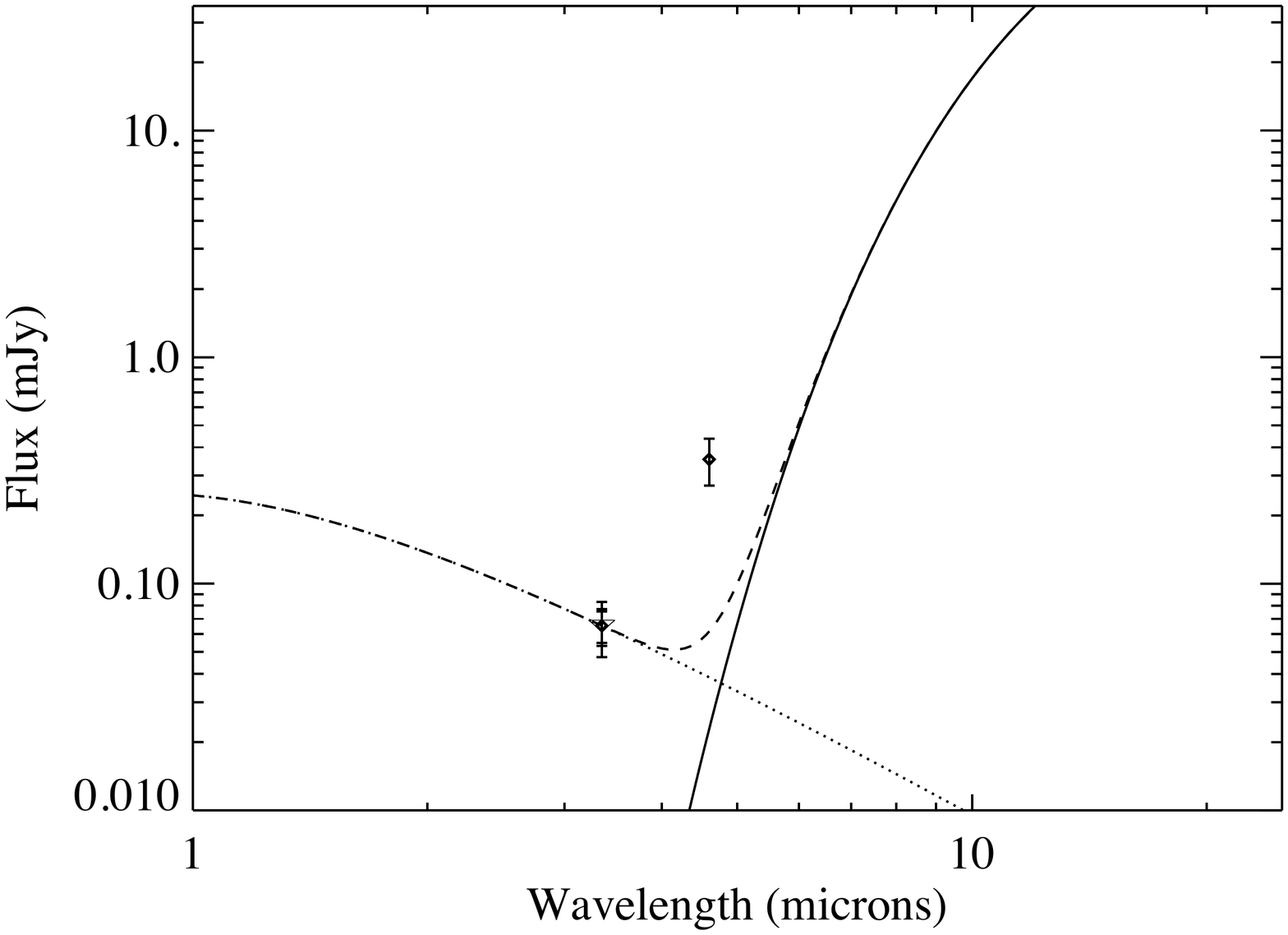}{0.44\textwidth}{(j)}
          }
\gridline{\fig{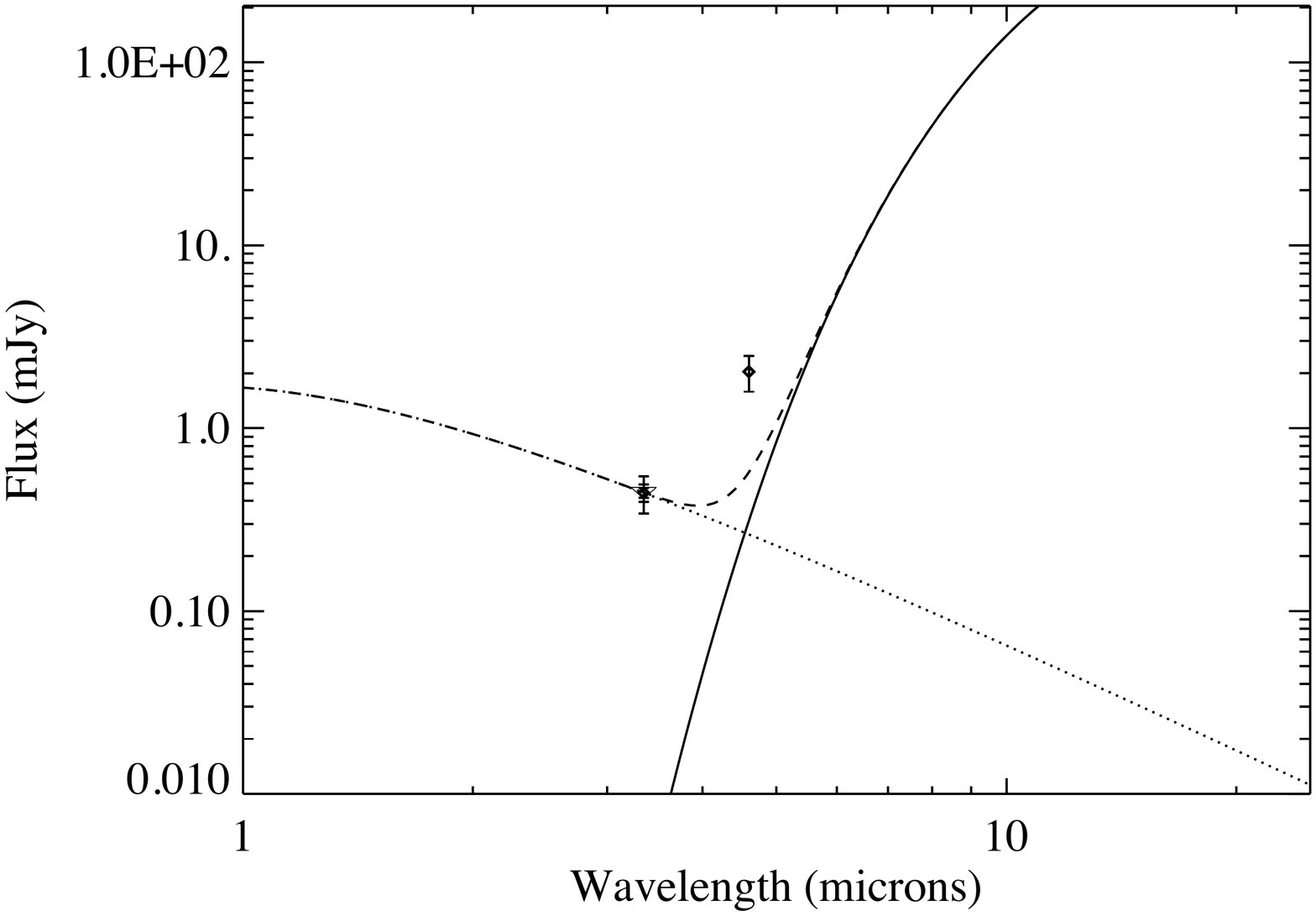}{0.44\textwidth}{(k)}
          \fig{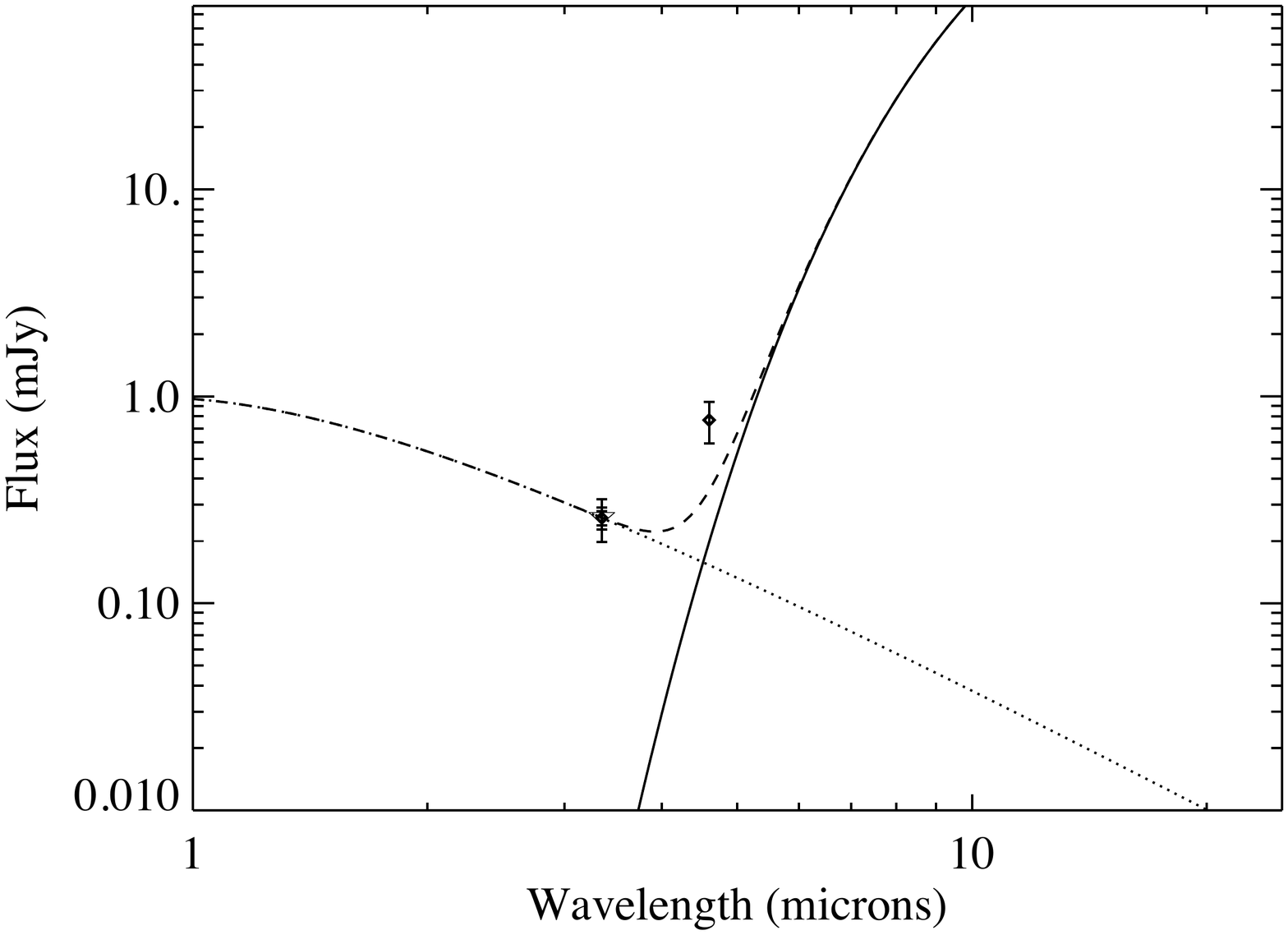}{0.44\textwidth}{(l)}
          }
\gridline{\fig{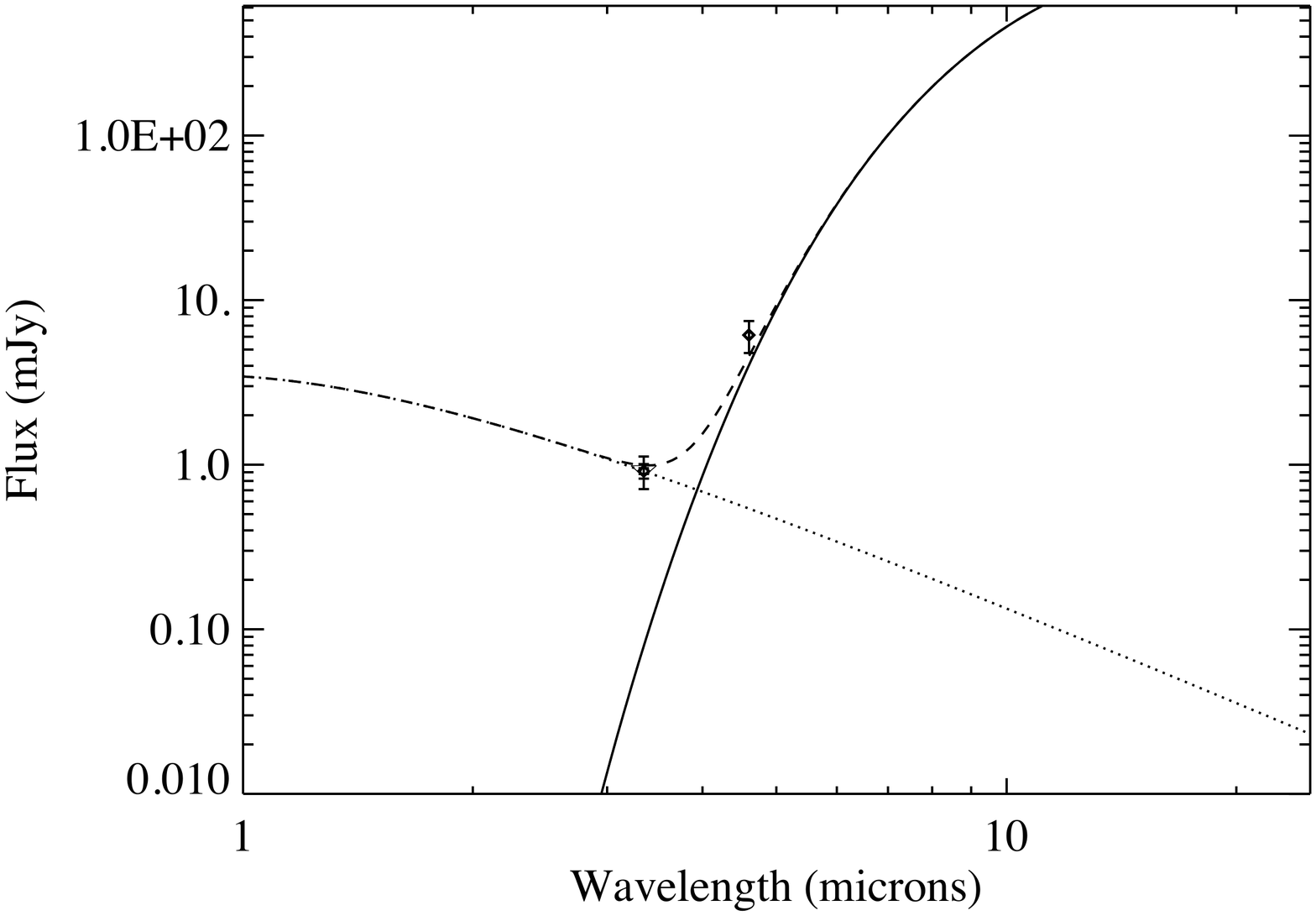}{0.44\textwidth}{(m)}
          \fig{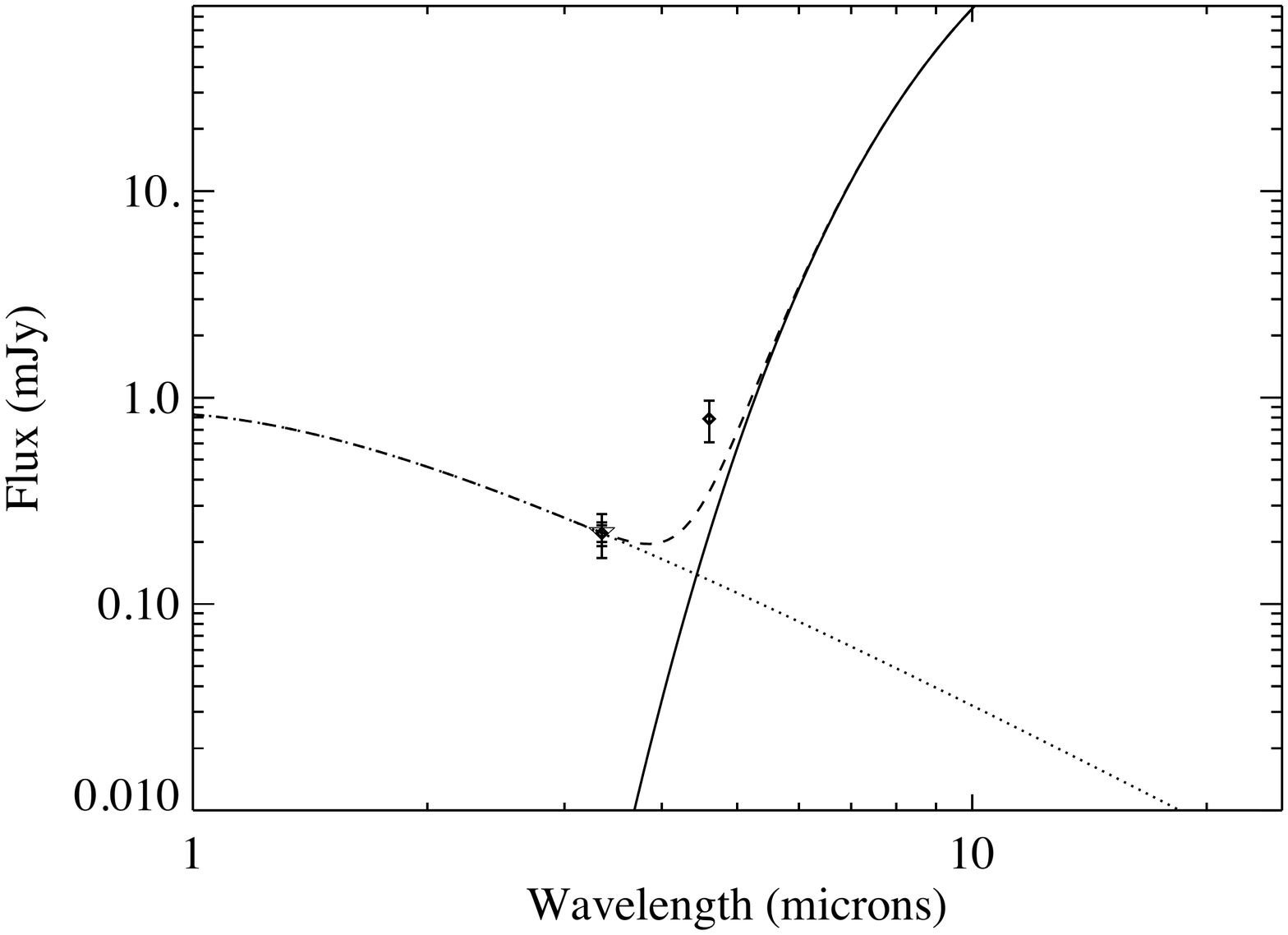}{0.44\textwidth}{(n)}
          }
\end{figure*}

\clearpage

\vfill
\begin{figure*}[h]

\gridline{\fig{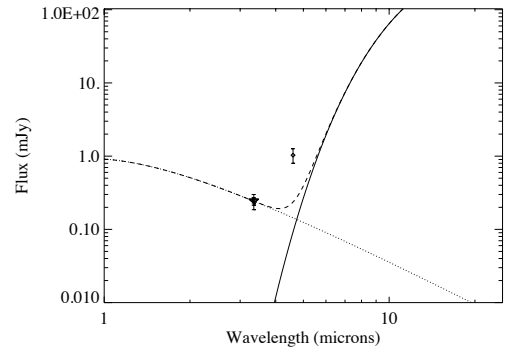}{0.44\textwidth}{(o)}
          }
\caption{Model of dust spectral flux distribution and 4.6 $\mu$m excess for HTCs with significant W2 excess (See Figure~\ref{DetectPlot}).\\
(a) C/2014 J1, (b) C/2014 Q3 Epoch 2, (c) C/2014 Q3 Epoch 3, (d) C/2014 W9 Epoch 2, (e) C/2014 W9 Epoch 3, (f) C/2015 A1, (g) C/2015 GX Epoch 3, (h) C/2015 GX Epoch 4, (i) C/2015 GX Epoch 5, (j) C/2015 H1 Epoch 4, (k) C/2015 X8 Epoch 1, (l) C/2015 X8 Epoch 2, (m) C/2015 YG1.\\
\label{fig:w2_excess_plots}}
\end{figure*}

\end{document}